\documentclass[aps,preprint,superscriptaddress,showpacs,floatfix]{revtex4-1}

\usepackage{colortbl}
\usepackage{graphicx}
\usepackage{dcolumn}
\usepackage{bm}
\usepackage{CJK}
\usepackage{amsmath,amssymb,latexsym}
\usepackage{multirow,array,booktabs,mathrsfs}

\begin{document}

\title{Octupole deformation and Ra puzzle in reflection asymmetric covariant density functional theory }

\author{L. F. Yu}
\affiliation{State Key Laboratory of Nuclear Physics and Technology, School of Physics, Peking University, Beijing 100871, China}%
\author{P. W. Zhao}
\affiliation{State Key Laboratory of Nuclear Physics and Technology, School of Physics, Peking University, Beijing 100871, China}%
\author{S. Q. Zhang}\email{sqzhang@pku.edu.cn}
\affiliation{State Key Laboratory of Nuclear Physics and Technology, School of Physics, Peking University, Beijing 100871, China}%
\author{J. Meng}\email{mengj@pku.edu.cn}
\affiliation{State Key Laboratory of Nuclear Physics and Technology, School of Physics, Peking University, Beijing 100871, China}%
\affiliation{School of Physics and Nuclear Energy Engineering, Beihang University, Beijing 100191, China}%
\affiliation{Department of Physics, University of Stellenbosch, Stellenbosch, South Africa}%

\begin{abstract}
Reflection asymmetric covariant density functional theory (CDFT) based on the point-coupling interaction is established on a two-center harmonic-oscillator basis and applied to investigate the Ra puzzle, i.e., the anomalous enhancement of the residual proton-neutron interactions $\delta V_{pn}$ for Ra isotopes around $N=135$.
The octupole deformation and shape evolution in the Ra and Rn isotopes are examined in the potential energy surfaces in ($\beta_2, \beta_3$) plane by the constrained reflection asymmetric calculations.
The $\delta V_{pn}$ values extracted from the double difference of the binding energies for Ra isotopes are compared with the data as well as the axial and the triaxial calculations. It is found that the octupole deformation is responsible for the Ra puzzle in the microscopic CDFT.
\end{abstract}
\pacs{21.10.Dr, 21.60.Jz, 27.80.+w, 27.90.+b }
\maketitle


\section{Introduction}

During the past several decades, the importance of the proton-neutron interaction in nuclear structure has been widely recognized~\cite{Talmi1962Rev.Mod.Phys.704}. It affects many aspects of nuclear structure, such as the
single-particle energy levels, the nuclear shape transition, the onset of deformation, the shell closure, etc.~\cite{Talmi1962Rev.Mod.Phys.704,Federman1977Phys.Lett.B385,Heyde1985Phys.Lett.B303,Zhang2005Nucl.Phys.A106}.

Since nuclear masses embody the interactions of all the nucleons, it is possible to isolate and extract the interaction of specific kind of nucleons from the masses. In particular, the average interaction strength $\delta V_{pn}$ between the last protons and the last neutrons in even-even and even-$Z$, odd-$N$ nuclei can be extracted by the double difference of the binding energies~\cite{Zhang1989Phys.Lett.B1,Zhao1995Nucl.Phys.A483} as
\begin{subequations}\label{eq-deltaVpn}
\begin{align}
       \delta V_{pn}^{ee}(Z, N)=&\frac{1}{4}\{[B(Z,N)-B(Z,N-2)]-\nonumber\\
       &[B(Z-2,N)-B(Z-2,N-2)]\},\\
       \delta V_{pn}^{eo}(Z, N)=&\frac{1}{2}\{[B(Z,N)-B(Z,N-1)]-\nonumber\\
       &[B(Z-2,N)-B(Z-2,N-1)]\}.
\end{align}
\end{subequations}
With the atomic mass evaluation published in 2003 (AME03)~\cite{Audi2003Nucl.Phys.A337}, a systematic investigation of $\delta V_{pn}$ values throughout the mass surface was performed in Refs.~\cite{Cakirli2005Phys.Rev.Lett.092501,Brenner2006Phys.Rev.C034315,Cakirli2006Phys.Rev.Lett.132501,Oktem2006Phys.Rev.C027304}. It was found that the results in regions of strong shell closures and in regions where shape transitions occur are especially interesting and are able to reflect structural features. Meanwhile, there are also many experimental~\cite{Hager2007Nucl.Phys.A20-39,Kellerbauer2007Phys.Rev.C045504,
Gomez-Hornillos2008Phys.Rev.C014311,Chen2009Phys.Rev.Lett.122503,Neidherr2009Phys.Rev.Lett.112501,
Breitenfeldt2010Phys.Rev.C034313,Ketelaer2011Phys.Rev.C014311} and theoretical~\cite{Stoitsov2007Phys.Rev.Lett.132502,Gao1999Phys.Rev.C735,Fu2010Phys.Rev.C034304} efforts which were devoted to investigate the $\delta V_{pn}$.

In particular, it was noted in Ref.~\cite{Brenner2006Phys.Rev.C034315} that there are anomalous enhancements of $\delta V_{pn}$ for $^{221}$Ra and $^{223}$Ra with $N=133,135$ deviating from the general trend of $\delta V_{pn}$ values, i.e., the so-called ``Ra puzzle''~\cite{Brenner2006Phys.Rev.C034315}.
Later on, a precise Penning-trap mass measurement on $^{223-229}$Rn has provided clear evidence of the existence of Ra puzzle and found that $\delta V_{pn}$ of the odd-$N$ Ra isotopes shows a well developed peak around $N=135$ which terminates at $N=139$~\cite{Neidherr2009Phys.Rev.Lett.112501}.

It is speculated that the Ra puzzle is associated with the softness of well-known octupole deformation in this region~\cite{Brenner2006Phys.Rev.C034315,Neidherr2009Phys.Rev.Lett.112501}. The octupole correlation is due to the  interaction between orbital pairs with $\Delta l=3$ and $\Delta j=3$ around the Fermi surface. For the nuclei around $Z=88$ and $N=134$, there exist octupole pairs ($\pi 2f_{7/2}$, $\pi 1i_{13/2}$) for protons and ($\nu 2g_{9/2}$, $\nu 1j_{15/2}$) for neutrons. Therefore, to understand the phenomenon of Ra puzzle, it is necessary to have a reliable theory including the reflection asymmetric degree of freedom.

The covariant density functional theory (CDFT) has achieved great successes in describing nuclear properties of both stable and exotic nuclei~\cite{Ring1996Prog.Part.Nucl.Phys.,Vretenar2005Phys.Rep.101,Meng2006Prog.Part.Nucl.Phys.}, including the recent achievements in nuclear
magnetic moments~\cite{Yao2006Phys.Rev.C024307,Li2011Prog.Theo.Phys.1185,Li2011Sci.ChinaPhys.Mech.Astron.204}, pseudospin symmetry~\cite{Zhou2003Phys.Rev.Lett.262501,Chen2003Chin.Phys.Lett.358,Ginocchio2005Phys.Rep.165,Liang2011Phys.Rev.C041301(R),Guo2012Phys.Rev.C021302,Lu2012Phys.Rev.Lett.072501},
low-lying excitations~\cite{Nikvsic2011Prog.Part.Nucl.Phys.519,Li2009Phys.Rev.C054301,Yao2011Phys.Rev.C014308}, magnetic and antimagnetic
rotations~\cite{Peng2008Phys.Rev.C,Zhao2011PhysicsLettersB181,Yu2012Phys.Rev.C024318,Zhao2011Phys.Rev.Lett.122501,Zhao2012Phys.Rev.C054310}, collective vibrations~\cite{N.Paar2007Rep.Prog.Phys.691,H.Z.Liang2008Phys.Rev.Lett.122502,Paar2009Phys.Rev.Lett.032502,Niu2009Phys.Lett.B315,Liang2009Phys.Rev.C064316}, and so on. Therefore, the CDFT with reflection asymmetry is an appropriate choice to investigate the Ra puzzle. In most of the successful versions of CDFT in nuclei, the Fock terms~\cite{Long2006Phys.Lett.B150} are not included explicitly, which leads to the relativistic mean-field (RMF) theory and forms the basis of its widespread applicability at
present.

The reflection asymmetric relativistic mean-field (RAS-RMF) theory with meson-exchange interaction has been independently developed in Ref.~\cite{Rutz1995Nucl.Phys.A680} on grid and in Ref.~\cite{Geng2007Chin.Phys.Lett.1865} on a two-center harmonic-oscillator (TCHO) basis~\cite{Greiner1994}.
This model has been successfully applied in the description of the ground-state properties of $^{226}$Ra~\cite{Geng2007Chin.Phys.Lett.1865} and the shape evolution of Sm~\cite{Zhang2010Phys.Rev.C034302} and Th~\cite{Guo2010Phys.Rev.C047301} isotopes.
In Ref.~\cite{Lu2012Phys.Rev.C011301,ZhaoarXiv:1209.6567v1[nucl-th]}, a RAS-RMF theory using both the meson-exchange and the point-coupling interactions with the triaxial degree of freedom was developed on the conventional harmonic-oscillator basis.

Recently, CDFT with the point-coupling interaction has attracted more and more attentions due to its simple applicability in being extended beyond the mean-field approximation~\cite{Nikolaus1992Phys.Rev.C1757,Burvenich2002Phys.Rev.C044308,Zhao2010Phys.Rev.C}. In this paper, the reflection asymmetric RMF theory with the point-coupling interaction (RAS-RMF-PC) is developed on a TCHO basis. With the RAS-RMF-PC thus implanted, the potential energy surfaces in ($\beta_2, \beta_3$) plane for Ra and Rn isotopes will be studied and the relationship between Ra puzzle and octupole deformation will be examined.

\section{THEORETICAL FRAMEWORK}%
\label{section1}
The starting point of the RAS-RMF-PC model is an effective Lagrangian density with the zero-range point-coupling interaction between nucleons.
By means of the conventional variation principle, the Dirac equation for nucleons can be obtained
\begin{equation}\label{Eq:Dirac-PC}
  [-i\bm\alpha\cdot\bm\nabla+\beta\gamma_\mu V^\mu+\beta(M+S)]\psi_k(\bm{r})=\varepsilon_k\psi_k(\bm{r}),
\end{equation}
where
\begin{equation}\label{Eq:potential-PC}
S(\bm{r})    =\alpha_S\rho_S+\beta_S\rho^2_S+\gamma_S\rho^3_S+\delta_S\triangle\rho_S,
\end{equation}
\begin{eqnarray}
V^\mu(\bm{r})&=&\alpha_Vj^\mu_V +\gamma_V (j^\mu_V)^3
                       +\delta_V\triangle j^\mu_V\nonumber\\
             & & +\tau_3\alpha_{TV} j^\mu_{TV}+ \tau_3\delta_{TV}\triangle j^\mu_{TV}+ e A^\mu.
\end{eqnarray}
More details can be found in Refs.~\cite{Nikolaus1992Phys.Rev.C1757,Burvenich2002Phys.Rev.C044308,Zhao2010Phys.Rev.C}. The basis expansion method is widely used to solve the Dirac equation. For nuclei with reflection asymmetry, the Dirac spinors could be expanded in terms of the eigenfunctions of the TCHO potential
\begin{equation}\label{tcpot}
    V(r_\bot, z)=\frac{1}{2}M\omega^2_\bot r^2_\bot+\left\{\begin{array}{c}
                                                           \frac{1}{2}M\omega^2_1(z+z_1)^2, \quad z<0, \\
                                                            \frac{1}{2}M\omega^2_2(z-z_2)^2, \quad z\geq0,
                                                            \end{array}
                                                    \right.
\end{equation}
where $z_1$ and $z_2$ (real, positive) are the distances between the centers of the spheroids and the intersection plane, and $\omega_1$ ($\omega_2$) are the corresponding oscillator frequencies for $z<0$ ($z\geq 0$)~\cite{Geng2007Chin.Phys.Lett.1865}. The TCHO basis can be completely specified by three parameters: $\delta_2$, $\delta_3$ and $\Delta z$, and their detailed definitions can be found in Ref.~\cite{Geng2007Chin.Phys.Lett.1865}.

The binding energy with a given deformation can be obtained by the deformation constrained calculation, i.e., by minimizing
\begin{equation}\label{eq:def-cons}
    \langle H'\rangle=\langle H\rangle +\frac{1}{2}C(\langle \hat{Q}_2\rangle-\mu_2)^2,
\end{equation}
where $C$ is a spring constant, $\mu_2$ is the given quadrupole moment, and $\langle \hat{Q}_2\rangle$ is the expectation value of qudrupole moment operator $\hat{Q}_2=2r^2P_2(\cos\theta)$. The octupole moment constraint can also be applied similarly with $\hat{Q}_3=2r^3P_3(\cos\theta)$. By constraining the quadrupole moment and octupole moment simultaneously, the total energies in ($\beta_2, \beta_3$) plane can be obtained.

\section{NUMERICAL DETAILS}
\label{section2}
In the present work, all the RMF calculations are performed with the newly proposed effective interaction  PC-PK1~\cite{Zhao2010Phys.Rev.C}. The pairing correlations are neglected at this moment. For the axial and triaxial calculations, the Dirac equations are solved on the three-dimensional isotropic harmonic-oscillator basis. For the RAS-RMF-PC calculations, the Dirac equations are solved on the TCHO basis with $\delta_2=0.0$, $\delta_3=0.99$ and $\Delta z\approx 0$. By increasing the major shell number of the harmonic-oscillator basis from $N_f=16$ to 18, the binding energy of $^{212}$Ra changes less than 0.01\% for axial and triaxial calculations, and less than 0.02\% for RAS-RMF-PC calculations. Therefore, the major shell number $N_f=16$ is adopted in the following calculations.

\section{RESULTS AND DISCUSSION}
\label{section3}
\begin{figure*}
\centerline{
\includegraphics[width=12cm]{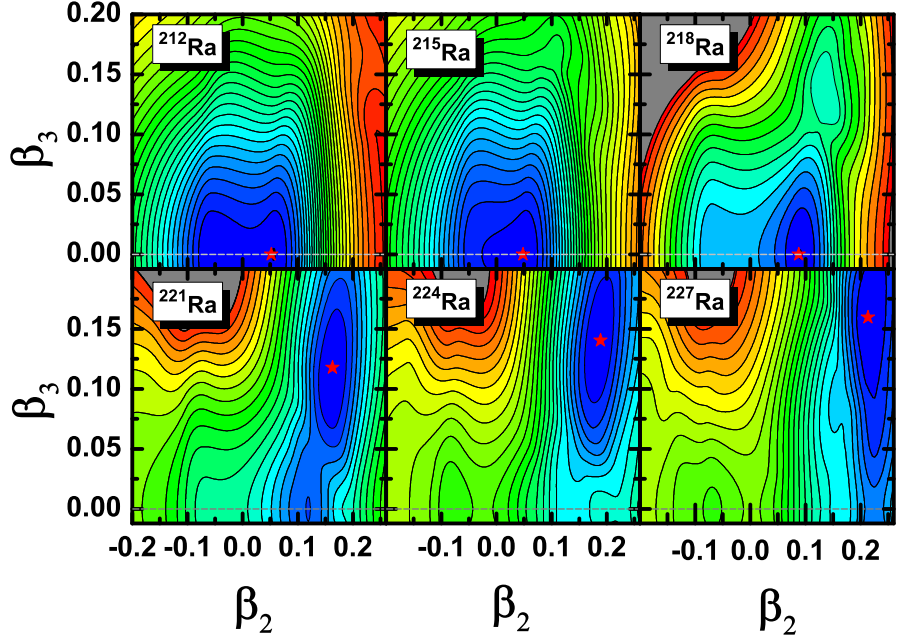}
}\caption{(Color online) The potential energy surfaces for Ra isotopes in
($\beta_2, \beta_3$) plane in the reflection asymmetric covariant density functional
calculations with PC-PK1. The energy difference between neighboring contour lines is 0.5 MeV.
The global minima are denoted by solid stars.}
\label{fig1}
\end{figure*}

In order to investigate the shape evolution in the Ra and Rn isotopes with the octupole degree of freedom, the potential energy surfaces in ($\beta_2, \beta_3$) plane for $^{210-229}$Ra and $^{208-227}$Rn have been calculated in the constrained RAS-RMF-PC theory. As examples, the contour plots for $^{212}$Ra, $^{215}$Ra, $^{218}$Ra, $^{221}$Ra, $^{224}$Ra, and $^{227}$Ra are shown in Fig.~\ref{fig1}, and the global minima therein are denoted by stars. It is shown that the ground states of $^{212}$Ra, $^{215}$Ra, and $^{218}$Ra are near spherical without octupole deformation. For $^{218}$Ra, apart from the global minimum with $\beta_3=0$, there exists a local minimum with the octupole deformation $\beta_3=0.13$. The octupole deformation appears in the ground states of $^{221}$Ra, $^{224}$Ra, and $^{227}$Ra, and increases with the neutron number.

By analyzing the potential energy surfaces for all the Ra isotopes, one could further see that the global minima for $^{210-219}$Ra are slightly prolate with $\beta_2<0.1$. Noted that for $^{217}$Ra, $^{218}$Ra, and $^{219}$Ra, there appears one local minimum with obvious octupole deformation. For $^{220-229}$Ra, the octupole deformed minimum becomes the global minimum, which means that the ground states of these isotopes are octupole deformed. Quantitatively, for $^{210-219}$Ra, the ground-state deformation parameters are of $0.04\leq\beta_2<0.1$ and $\beta_3=0$. In particular, the ground state of the magic nucleus $^{214}$Ra is near spherical ($\beta_2=0.04$). For the ground states of $^{220-229}$Ra, both the quadrupole and octupole deformation parameters become remarkable with $0.15<\beta_2<0.23$ and $0.1\leq\beta_3\leq 0.18$, and increase with the neutron number.

The shape evolution of Rn isotopes is similar to that of Ra isotopes. The ground states of $^{208-217}$Rn have only quadrupole deformation and the $\beta_2$ varies in the range from -0.07 to 0.11. The octupole deformation appears in the ground states of $^{218-225}$Rn, and the $\beta_3$ increases from 0.02 for $^{218}$Rn to 0.08 for $^{223}$Rn but finally drop to 0.02 for $^{225}$Rn. The octupole deformation further vanishes in the ground states of $^{226}$Rn and $^{227}$Rn. Compared to the Ra isotopes,  the Rn isotopes with the same neutron number have smaller octupole deformations.

\begin{figure}
\centerline{
\includegraphics[width=8cm]{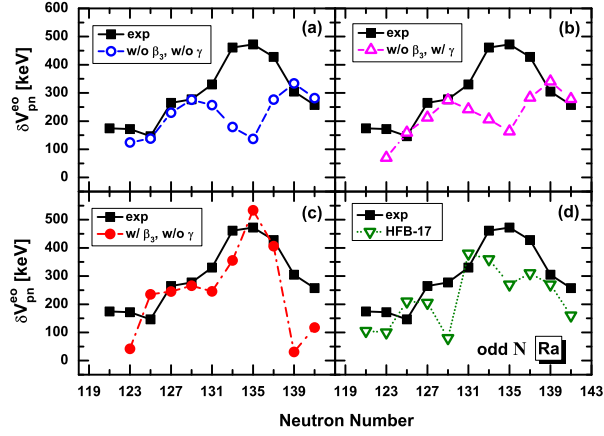}
} \caption{(Color online) The residual proton-neutron interactions $\delta V_{pn}$ data (solid squares)~\cite{Neidherr2009Phys.Rev.Lett.112501} for odd-$N$ Ra isotopes in comparison
with the calculated values by axial (a), triaxial (b), and octupole (c) CDFT with PC-PK1,
as well as HFB-17 mass table (d)~\cite{Goriely2009Phys.Rev.Lett.152503}.}
\label{fig2}
\end{figure}

\begin{figure}
\centerline{
\includegraphics[width=8cm]{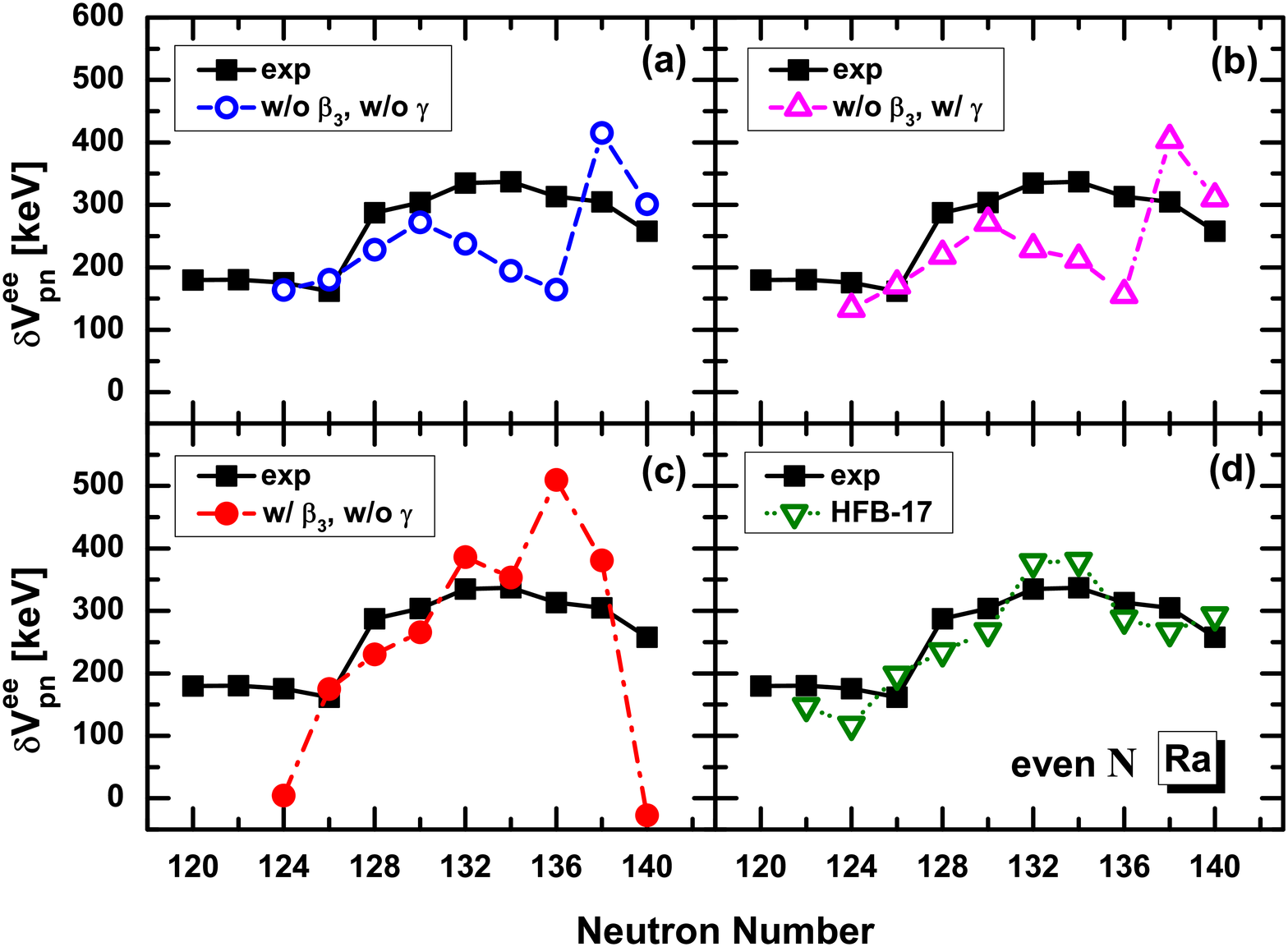}
} \caption{(Color online) Same as Fig.~\ref{fig2}, but for the even-even Ra isotopes.}
\label{fig3}
\end{figure}

From the binding energies of neighboring nuclei, the residual proton-neutron interactions $\delta V_{pn}$ for Ra isotopes can be extracted according to Eq.(\ref{eq-deltaVpn}). In Fig.~\ref{fig2}, the $\delta V_{pn}$ values extracted from RMF models are compared with the empirical values from the data~\cite{Neidherr2009Phys.Rev.Lett.112501} for odd-$N$ Ra isotopes. Here, the results in the axial and the triaxial as well as the reflection asymmetric RMF calculations are respectively denoted by open circles, triangles, and solid circles.

In Fig.~\ref{fig2} (a), the axial RMF calculations reproduce the data well except the data for $^{221}$Ra, $^{223}$Ra, and $^{225}$Ra, and thus fails in reproducing the peak around $N=135$. The same conclusion remains even after the triaxiality is considered, as shown in Fig.~\ref{fig2}(b). This indicates that the quadrupole deformation and triaxiality are not the reasons for the Ra puzzle.

After including the octupole degree of freedom, as shown in Fig.~\ref{fig2} (c), the peak around $N=135$ for the $\delta V_{pn}$ value is well reproduced. This clearly indicates that the Ra puzzle can be well understood with the octupole deformation. It should be pointed out that the discrepancies appear at $^{227}\rm Ra$ and $^{229}\rm Ra$, which might be attributed to the pairing correlation neglected in present RAS-RMF calculations.

For comparison, the results from the nuclear mass tables HFB-17~\cite{Goriely2009Phys.Rev.Lett.152503} are presented in Fig.~\ref{fig2} (d) and compared with the data. Again, the HFB-17 results fail to reproduce the peak around $N=135$ due to the absence of the octupole degree of freedom.

Similar to Fig.~\ref{fig2}, the residual proton-neutron interactions $\delta V_{pn}$ data for even-even Ra isotopes are shown in Fig.~\ref{fig3} in comparison with the corresponding calculations. The data for light Ra isotopes are well reproduced in both the axial and triaxial calculations. However, the discrepancies for $\delta V_{pn}$ appear for the Ra isotopes from $N=132$ to $136$. After including the octupole deformation, the data are well reproduced expect for $^{224}$Ra and $^{228}$Ra.

Comparing with Fig.~\ref{fig2}(c), the agreement with the data for the even-even Ra isotopes is less
impressive than that for the odd-$N$ isotopes. Further improvement may be achieved by taking into account the pairing correlation.
In fact, as shown in Fig.~\ref{fig3}(d), the results from HFB-17 mass table with the pairing correlation could reproduce the data quite well.
Therefore, in the relativistic framework, both the octupole deformation and the pairing correlation might be important to describe the $\delta V_{pn}$ values for the even-even Ra isotopes.

\begin{figure*}
\centerline{
\includegraphics[width=12cm]{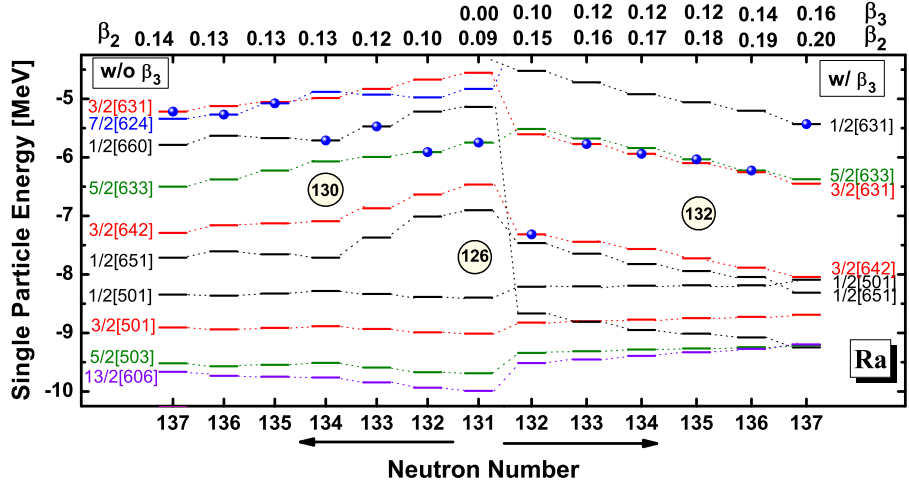}
}
\caption{(Color online) Neutron single-particle levels for $^{219-225}$Ra in the RS-RMF-PC (left) and RAS-RMF-PC (right) calculations. The solid circles denote the levels occupied by the last neutron. The levels are labeled by Nilsson-like notations $\Omega[Nn_zm_l]$ of the first component. On the top margin are the deformations. }
\label{fig4}
\end{figure*}


For odd-$N$ isotopes, the $\delta V_{pn}$ in Eq.~\ref{eq-deltaVpn}(b) could be rewritten as
\begin{equation}
\delta V_{pn}^{eo}(Z, N)=\frac{1}{2}[S_n(Z,N)-S_n(Z-2,N)],
\end{equation}
with the single-neutron separation energy $S_n(Z,N)=B(Z,N)-B(Z,N-1)$.
For odd-$N$ Ra isotopes, they are the single-neutron separation energy differences between the Ra isotopes and the Rn isotopes. In the RAS-RMF calculations, for Rn isotopes, as their octupole deformations are quite small, the $S_n$ calculated with and without octupole degree of freedom is similar.
For Ra isotopes, as their octupole deformations are remarkable, the $S_n$ calculated with and without octupole degree of freedom is considerably different. Therefore, the difference of $\delta V_{pn}$ values with or without octupole deformation is determined by the corresponding difference of the single-neutron separation energy for Ra isotopes.

As the single-neutron separation energy is nothing but the Fermi surface in the single-particle spectrum with pairing neglected, it is interesting to examine the effect of the octupole deformation on the neutron single-particle levels,
as shown in Fig.~\ref{fig4}, for $^{219-225}$Ra obtained by the RMF calculations with and without reflection symmetry. For $^{219}$Ra, the same single-particle level structure is obtained as $\beta_3=0$.
From $^{220}$Ra to $^{225}$Ra, the level structure is quite different due to the octupole deformation.
In the axial calculations, the $\beta_2$ deformation increases with the neutron number and the shell gap at magic number $N=126$ in $^{219}$Ra gradually disappears. In RAS-RMF calculations, however, due to the performance of octupole deformation the shell gap at $N = 126$  disappears and a large energy gap at $N = 132$ appears for $^{220-225}$Ra.

By switching on the octupole deformation, the Fermi energies of $^{220-225}$Ra are bound more deeply, which results in   larger single-neutron separation energies and provides reasonable explanation for the Ra puzzle, as shown in Fig.~\ref{fig2}(c).

\section{Conclusion}
\label{conclusion}

In summary, the reflection asymmetric CDFT based on the point-coupling interaction is established on a two-center harmonic-oscillator basis. The potential energy surfaces in ($\beta_2, \beta_3$) plane are calculated by the constrained reflection asymmetric calculations to investigate the shape evolution in the Ra and Rn isotopes. It is found that the ground states are near-spherical for $^{210-219}$Ra and have remarkable octupole deformation for $^{220-229}$Ra. In comparison, the Rn isotopes have smaller octupole deformations but similar shape evolutive behavior as the Ra isotopes.

The residual proton-neutron interactions $\delta V_{pn}$ for Ra isotopes are extracted from the double difference of the binding energies of Ra and Rn isotopes and are compared with the experimental values as well as the axial and triaxial RMF calculations. It is found that the octupole deformation provides a reasonable explanation for the Ra puzzle, i.e., the anomalous enhancement of $\delta V_{pn}$ for Ra isotopes around $N=135$.

This explanation for the Ra puzzle by the octupole deformation can be traced back to the single-neutron separation energy and the single-particle energy spectrum. The octupole deformation will drive the Fermi surface to be bound more deeply for $^{220-225}$Ra, which results in larger single-neutron separation energy as well as the appearance of an energy gap at $N=132$ for $^{220-225}$Ra.

{\center{\bf ACKNOWLEDGMENTS}}

We are grateful to B. N. Lu and S. G. Zhou for helpful discussions. This work was partly supported by the Major State 973 Program under Grants No. 2013CB834400, the National Natural Science Foundation of China
under Grants No. 10975007, No. 10975008, No. 11105005, and No. 11175002, the Research Fund for the Doctoral Program of Higher Education under Grant No. 20110001110087.



%
%

\end{document}